\documentclass{article}
\usepackage{ijcai23}

\usepackage{times}
\usepackage{soul}
\usepackage{url}
\usepackage[hidelinks]{hyperref}
\usepackage[utf8]{inputenc}
\usepackage[small]{caption}
\usepackage{graphicx}
\usepackage{amsmath}
\usepackage{amsthm}
\usepackage{booktabs}
\usepackage{algorithm}
\usepackage{algorithmic}
\usepackage{multirow}
\usepackage{booktabs}
\usepackage{tabularx}
\usepackage[switch]{lineno}

\title{Individually Rational Collaborative Vehicle Routing\\ through Give-And-Take Exchanges}

\author{
    Paul Mingzheng Tang, Ba Phong Tran and Hoong Chuin Lau
    \affiliations
    School of Computing and Information Systems, Singapore Management University
    \emails
    \{paultang, bptran, hclau\}@smu.edu.sg
}

\begin{document}

\maketitle

\begin{abstract}
In this paper, we are concerned with the automated exchange of orders between logistics companies in a marketplace platform to optimize total revenues. We introduce a novel multi-agent approach to this problem, focusing on the Collaborative Vehicle Routing Problem (CVRP) through the lens of individual rationality. Our proposed algorithm applies the principles of Vehicle Routing Problem (VRP) to pairs of vehicles from different logistics companies, optimizing the overall routes while considering standard VRP constraints plus individual rationality constraints. By facilitating cooperation among competing logistics agents through a Give-and-Take approach, we show that it is possible to reduce travel distance and increase operational efficiency system-wide. More importantly, our approach ensures individual rationality and faster convergence, which are important properties of ensuring the long-term sustainability of the marketplace platform. We demonstrate the efficacy of our approach through extensive experiments using real-world test data from major logistics companies. The results reveal our algorithm's ability to rapidly identify numerous optimal solutions, underscoring its practical applicability and potential to transform the logistics industry.
\footnote{This paper was presented in the IJCAI 2023 First International Workshop on Search and Planning with Complex Objectives (WoSePCO) http://idm-lab.org/wiki/complex-objective.}
\end{abstract}

\section{Introduction}
The logistics sector is facing a rapidly evolving landscape, with the surge of e-commerce and the escalating demand for efficient supply chain management propelling its growth. In this dynamic operational environment, logistics service providers (LSPs) are in constant pursuit of innovative strategies to enhance their operational performance, reduce costs, and maintain a competitive edge. One such strategy that has gained prominence is the collaboration among multiple LSPs. This collaborative approach allows LSPs to share jobs with their competitors, thereby optimizing overall profits while ensuring individual profitability. Moreover, a collaborative platform that enhances routing efficiency offers not only economic benefits to the LSPs but also contributes to environmental sustainability by reducing carbon emissions and alleviating road congestion.

In this paper, we delve into this complex planning problem from the perspective of a central platform that serves as a marketplace for LSPs to share jobs. This platform presents a unique challenge: it must maximize the total efficiency of routes while ensuring that the solutions are individually rational, i.e., each LSP has the incentive to participate in the marketplace. In multiagent terminology,  participation in a mechanism is individually rational to an agent if the agent's payoff in participation is no less than the payoff that the agent would get by not participating in it.

More specifically, this work is to address a generalized variant of the Vehicle Routing Problem (VRP), specifically the Pickup and Delivery Problem with Time Windows (PDPTW), and extend it to a multi-LSP context that necessitates individually rational solutions. Our approach builds upon the \textit{Order Package Heuristic} proposed by \cite{de_jonge_heuristic_2021}, but with significant enhancements to address the limitations of their approach.

The aforementioned approach \cite{de_jonge_heuristic_2021} has some limitations. The algorithm is based on the exchange of orders between vehicles, which assumes that companies only negotiate about which company will deliver which orders, and not about any form of financial compensation for the delivery of another company's orders. This assumption simplifies the problem into a single-objective optimization problem, which is not very interesting compared to the state-of-the-art. Furthermore, their partners have indicated that automated price negotiations are not acceptable in a true working system. They require prices to be fixed over a longer term, such as a whole year. Automated day-to-day price negotiations would yield an opaque pricing mechanism with possibly highly fluctuating prices, which would be a serious problem for their bookkeeping.

In contrast, our approach aims to address these limitations by proposing a new algorithm that optimizes the routes for multiple vehicles owned by multiple LSPs. This approach not only maintains individual rationality property but also ensures faster convergence, while simultaneously optimizing vehicle scheduling to enhance overall efficiency. We demonstrate the efficacy of our algorithm through extensive experiments using real-world test data from major logistics companies. The results reveal that our approach can rapidly identify optimal solutions, highlighting its practical applicability and potential to be adopted by the logistics industry.

Our paper makes the following contributions:
\begin{itemize}
    \item A new heuristic that is competitive with existing approaches that generate solutions based on automatic negotiation. Furthermore, the solutions generated satisfy the individually rational properties.
    \item An analysis of the limitations of existing approaches with toy examples.
    \item Experiments on real-world and synthesized data that show competitive performance against the existing approaches.
\end{itemize}

\section{Related Works}
\subsection{Collaborative Vehicle Routing Problems}
A survey by \cite{gansterer_collaborative_2017} classifies methods to solve collaborative vehicle routing problems into 3 categories: centralized collaborative planning, auction-based decentralized planning, and decentralized planning without auctions. 
In terms of auction-based collaborative vehicle routing, \cite{zhang2019exact} proposed a branch-and-price-and-cut algorithm and a branch-and-bound algorithm to solve the vehicle routing problem with time windows and combinatorial auction. \cite{HAMMAMI2019150} addressed a bid construction problem with a heterogeneous fleet, and solved by exact and heuristic approaches based on adaptive large neighborhood search. They showed that an exact solver (CPLEX) could not obtain solutions better than  their best heuristic solution provided as an initial solution within certain time limits. More recently, \cite{Lau2021} considered an approach associated with a marketplace platform where the LSPs are expected to bid for job bundles while the auctioneer solves the Winner Determination Problem for this combinatorial auction via an exact and heuristic approach. 

Similar to \cite{de_jonge_multi-objective_2022} we focus on the decentralized planning without auctions applied to the generalized VRP variant, Pickup and Delivery Problem with Time Windows (PDPTW). Our problem statement follows closely with \cite{wang_operational_2014} and \cite{wang_rolling_2015}, where we maximize the total profit, however for our problem we must also maintain individual rationality while not using an auction-based system. Because our work builds upon \cite{de_jonge_multi-objective_2022}, our method can extend beyond our current problem where we are acting as the collaborative platform but also be used from the LSP's point of view to be used for automatic negotiation. 

\subsection{Multi-Objective VRP}
The Multi-objective Vehicle Routing Problem (MOVRP) extends the VRP by considering multiple objectives simultaneously, such as minimizing cost, minimizing time, and maximizing customer satisfaction. This multi-objective approach provides a more realistic model for real-world logistics operations, where decision-makers often need to balance conflicting objectives.

Several studies have proposed different methods for solving the MOVRP. \cite{molina2014multi} proposed a multi-objective model based on Tchebycheff methods for VRP with a heterogeneous fleet. However, the reliance on Tchebycheff's methods may limit the model's effectiveness in handling complex, real-world scenarios where multiple objectives are not easily separable or quantifiable. 

Kumar et al. \cite{kumar2014solving} introduced a Fitness Aggregated Genetic Algorithm (FAGA) for solving the multi-objective problem. While genetic algorithms can be powerful optimization tools, they can also be computationally expensive and may struggle with problems that require a balance between exploration and exploitation. The author \cite{miranda2018algorithms} tackled a multi-objective vehicle routing problem with hard time windows and stochastic travel and service times. They proposed two algorithms (a Multi-Objective Memetic Algorithm and a Multi-Objective Iterated Local Search) and compared them to an evolutionary multi-objective optimizer from the literature. However, the effectiveness of these algorithms in different scenarios and datasets remains to be fully explored.

\cite{defryn_multi-objective_2018} considers optimizing each LSP separately using a partner efficiency approach where each LSP's objective is optimized individually. While this method finds Pareto-optimal solutions, it does not guarantee individual rationality for all LSPs. Unlike previous works, our paper utilizes different multi-objective for each LSP company thus covering diverse scenarios. 

\section{Market Place}
The Market Place is an emerging business model where each agent operates to maximize its payoff, yet cooperates in the marketplace with other agents to exchange jobs to further optimize operational costs. It functions as a virtual meeting point for LSPs, allowing them to collaborate and optimize their operations while maintaining their individual interests. The marketplace aims to improve capacity utilization, reduce empty miles, streamline processes, and ultimately reduce costs for all participants. 

The input of the marketplace comprises the jobs for individual LSPs. Each job contains information such as timing, capacity, pickup, and delivery window. Furthermore, it requires to have a price that the LSP would gain after delivering this job. If there is no price associated, we need to develop an algorithm to determine this job's price based on the myriad factors such as the spatial and temporal component of the job related to the rest of the jobs to which it initially belonged (i.e., is this job isolated from the rest of the jobs on the same fleet?) There is no need to know the cost structure of each LSP as we assume that the cost is fixed using a predefined formula that is the same across the LSPS joining the marketplace. However, flexibility exists for other LSPs to join and propose their own objective to fit their own interests. 

The goal of the marketplace is to generate a set of orders to recommend to each LSP. Ideally, each LSP increases its efficiency by adopting the set of recommended orders. The concept is similar to automated negotiation \cite{de_jonge_multi-objective_2022}, in which each agent is purely self-interested and proposes solutions to other competitors yet remains cooperative to find mutual benefits. In our case, we have a centralized platform agent to perform the task of automated negotiations among LSP agents in the sense that it considers individual benefits and optimizes mutual benefits such that each agent wins by participating in the platform (vs not participating). To find which jobs to exchange which one in which agents, we employ VRP to solve the vehicle-pair exchanges. This approach ensures that the marketplace's recommendations are not only beneficial to individual LSPs but also optimize the overall efficiency of the logistics network.

\section{Problem Definition}
The following definitions describe our problem statement formally.

\textbf{Definition 1}: A \textbf{time-distance matrix} $(T,D)$, where $T$ and $D$ are both matrices $|L|\times|L|$. $L$ is the set of locations and $|L|$ represents the number of locations. The time and distance it takes to travel from location $A$ to location $B$ is denoted by $T(A,B)$ and $D(A,B)$ respectively. In this paper, a location consists of a latitude and longitude, but is not restricted to be so.

\textbf{Definition 2}: A \textbf{waypoint} is a tuple $(loc, st, et, service, vol, order)$. $loc$ is the location of the waypoint. $st$ and $et$ are the start and end times to reach the waypoint. $service$ is the service time incurred when the vehicle finishes processing the waypoint along its route. $vol$ is the change in capacity when a vehicle processes the waypoint along its route. $order$ is the order for this waypoint.

\textbf{Definition 3}: An \textbf{order} is a tuple $(P,D,rev)$. $P$ and $D$ are the pickup and drop-off waypoints. $rev$ is the revenue earned for processing the order.

\textbf{Definition 4}: A \textbf{vehicle} is a tuple $(cap, lspid, depot)$. $cap$ is the maximum capacity the vehicle can have at any point in its schedule. $lspid$ is the ID of the LSP the vehicle belongs to. $depot$ is the start and end waypoint required for the vehicle's assigned schedule.

\textbf{Definition 5}: A \textbf{vehicle schedule} is an ordered list of waypoints $W = (w_{1}, w_{2}, ..., w_{k})$ where $k$ is the number of waypoints in the route, and a vehicle assigned to process these waypoints. The first and last waypoints must be the vehicle's depot. The waypoints in the vehicle schedule must contain both the pickup and drop-off waypoints $P$ and $D$ if it contains the order, and the list must preserve the precedence constraint where $P$ is processed before $D$. The vehicle time schedule can also be inferred from the list of waypoints and consists of a tuple $t_{1}, t_{2}, ..., t{k}$. The vehicle time schedule is obtained by computing the time processed between waypoints and calculating the cumulative sum of $T(w_{i}.loc, w_{i+1}.loc) + w_{i}.service$. It maintains the time window constraint where $w_{i}.st \le t_{i} \le w_{i}.et$. Lastly, the cumulative load at any point of the vehicle schedule is less than the maximum capacity of the vehicle: $\sum_{i=1}^{j}{w_{i}.vol} \le V.cap$. The set of orders $G = (w_{1}.orderid, ..., w_{k}.orderid)$ can be inferred.

\textbf{Definition 6}: An \textbf{LSP} is represented by a fleet of vehicles $F = (V_1, V_2, ... V_m)$ where $m$ is the number of vehicles available for the LSP, and the cost function parameters $\alpha$ (cost per kilometer traveled) and $\beta$ (fixed cost of a vehicle if used). Hence, the cost of a vehicle would be: 

\[Cost(V) = \beta + \alpha \sum_{i}^{k-1}{D(V.w_{i}.loc, V.w_{i+1}.loc)}\].

The revenue of a vehicle would be the sum of the revenue of orders in each vehicle's schedule.

\[Rev(V) = \sum_{o \in G}{o.rev}\]

The final solution must maintain the Individual Rationality constraint, where the final profits earned must be greater than $Init$ which would be the profit earned by the LSP if it were to route without the marketplace platform.

\[\sum_{i}^{m} Rev(V_{i}) - Cost(V_{i}) \ge Init\]

\textbf{Problem Statement}: The goal is to distribute the initial set or orders $G_i$ shared by each LSP $i$ such that the total profit is maximized while maintaining individual rationality (IR) property. Hence, the expected output would be an updated set of order $G'_i$ for each LSP (along with the routes of the vehicles it owns).

\section{Give-And-Take (GAT) Heuristic}
As mentioned earlier, our heuristic is built upon \cite{de_jonge_heuristic_2021}, which uses the \textit{Order Package Heuristic}(OPH) to find solutions to the problem. An ordered package is defined by \cite{de_jonge_heuristic_2021} as a subset of orders in a schedule that corresponds to a sequence of consecutive locations.

Before delving into our heuristic, we can analyze the limitations of OPH which explains the motivation behind our method. To summarise OPH, the procedure is as follows:

\begin{enumerate}
    \item Find compatible order-vehicle pairs
    \item Determine all order packages
    \item Generate one-to-one exchanges (possible assignments of order packages to vehicles)
    \item Combine one-to-one exchanges into full exchanges (a possible subset of assignments in the previous step)
\end{enumerate}

When generating full exchanges, it is important to note that the change in cost for an assignment is computed with the assumption that the vehicle does not perform other assignments associated with that vehicle. To guarantee feasibility, a vehicle needs to be assigned to be either a donating or a receiving vehicle, and a receiving vehicle can only allow 1 assignment. This implies that a single iteration of OPH will not be able to find a solution where a pair of vehicles can both donate and receive orders from each other. The toy example shown in \autoref{fig:toy} illustrates this limitation. With this in mind, we propose a new heuristic algorithm called Give-And-Take (GAT) heuristic that aims to overcome this.

The idea is to solve a 2-vehicle VRP for every pair of vehicle schedules in the initial solution. We then also consider after solving the 2-vehicle VRP whether the vehicle schedules obtained can be swapped. We describe our procedure in detail in the rest of this section.

\subsection{Step 1: Solve 2-vehicle VRP for every pair}
We take every pair of vehicles from the current solution (For the first iteration, the current solution is obtained by solving the VRP for each LSP's set of orders without collaboration) and solve the 2-vehicle VRP for the orders in both vehicles' schedules. For every pair $(V_i,V_j)$, we run a VRP solver $VRP(V_{i}.G + V_{j}.G, vehicles=2)$ and obtain 2 updated vehicle schedules $(V'_i, V'_j)$. Take note that the vehicles can come from the same LSP, allowing for further improvements when rerunning GAT. Each VRP solved constitutes a possible action $A_{i,j}$ in which $(V_i, V_j)$ is updated to $(V'_i,V'_j)$, and added to the set of all possible actions $A$. The VRP solver minimizes cost and does not maintain individual rationality.

\subsection{Step 2: Add swap for each 2-vehicle VRP solution}
We expand the search space by also considering if we can swap all the waypoints between $V'_i$ and $V'_j$, and this action is denoted by $A'_{i,j}$. Hence, each solution produced when solving a 2-vehicle VRP will generate 2 possible actions, $A_{i,j}$ and $A'_{i,j}$. We can prune the number of actions if $A_{i,j}$ does not modify the original schedules.

The pseudo-code for Step 1 and 2 is shown in Algorithm \autoref{algorithm:gen_action_algorithm}.

\label{algorithm:gen_action_algorithm}
\begin{algorithm}[tb]
    \caption{Generate Actions}
    \textbf{Input}: List of vehicles $V$\\
    \textbf{Parameter}: - \\
    \textbf{Output}: List of actions $A$
    \begin{algorithmic}[1] 
        \STATE Let $A$ be an empty list.
        \STATE Let $N = |V|$
        \FOR{$i \in [1,N]$}
            \FOR{$j \in [i+1,N]$}
                \STATE Let $G = V_i.Orders + V_j.Orders$
                \STATE Let $(V'i, V'j)$ = VRP($G$, NumVehicles=2)
                \STATE Let $A_{i,j} = (V'i,V'j)$
                \IF{$(V'i, V'j) \neq (Vi, Vj)$}
                    \STATE Append $A_{i,j}$ to $A$
                \ENDIF
                \STATE $A'_{i,j} = (V'j,V'i)$
                \STATE Append $A'_{i,j}$ to $A$
            \ENDFOR
        \ENDFOR
        \STATE \textbf{return} $A$
    \end{algorithmic}
\end{algorithm}

\subsection{Step 3: Combine possible exchanges}
Similar to the combination mechanism used for OPH in \cite{de_jonge_heuristic_2021}, we formulate the problem as an ordinary constraint optimization problem with linear objective functions. The constraints imposed differ slightly, where every vehicle in our formulation can only have at most 1 action assigned, and the individual rationality constraints are still imposed. 

Because we are solving from the perspective of a central server, we only aim to minimize total costs to all LSPs while maintaining individual rationality, and hence we do not require a set of possible solutions (although it is possible to generate). Instead, we find a single best solution by passing the set of actions $A$ and the change in profits for each action for each LSP to a constraint optimization problem solver (Google ORTools). The authors in \cite{de_jonge_heuristic_2021} use their own multi-objective optimisation problem solver to obtain a Pareto frontier of solutions for automatic negotiation.

An important thing to note is that since the updated vehicle schedules have been computed in the previous steps already for GAT, we do not need to recompute the schedules after choosing the best subset of actions to perform. In contrast, OPH allows a vehicle to donate more than 1 order package per iteration and hence computing the schedule is required after an iteration of OPH. Additionally, as mentioned in \cite{de_jonge_heuristic_2021}, OPH can underestimate the cost savings when donating multiple order packages, while GAT obtains an exact change in cost and revenue for any combination found.

\subsection{Step 4: Rerun Heuristic}
Unlike the work in \cite{de_jonge_heuristic_2021}, we rerun steps 1 to 3 over multiple iterations to find better solutions. Rerunning OPH shows strong performance in our experiments, however, the time taken to run further iterations is much longer than anticipated. By using a fast local search heuristic for our VRP solver, we can run many iterations of GAT and obtain competitive performance.

The final set of orders $G'_i$ for each LSP $i$ can be obtained by the final schedules in each LSP's fleet.

\section{Experiments}
\subsection{Toy Sample}
\begin{figure} [!htbp]	
\centering
	\includegraphics[width=0.45\textwidth]{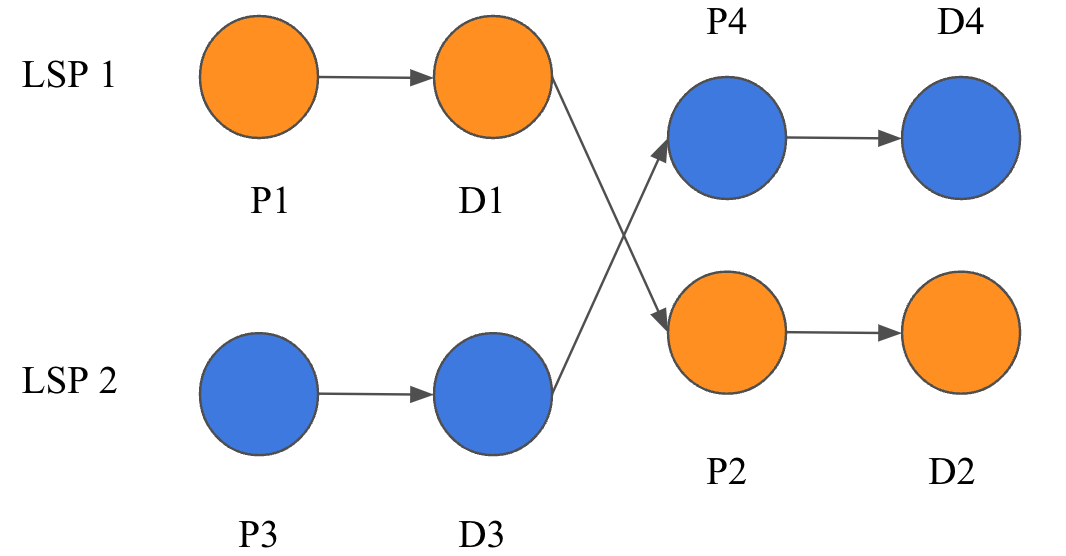}
	\caption{An illustration of GAT over OPH}
	\label{fig:toy}
\end{figure}
The toy example shown in \autoref{fig:toy} illustrates the limitations of the Order Package Heuristic (OPH) method in \cite{de_jonge_heuristic_2021}. As shown in \autoref{fig:toy}, let $V_1$ belonging to LSP 1 have the schedule $(P_1, D_1, P_2, D_2)$ and 
$V_2$ belonging to LSP 2 have the schedule $(P_3, D_3, P_4, D_4)$. The optimal solution is for $V_1$ and $V_2$ to swap orders 2 ($P_2,D_2$) and 4 ($P_4,D_4$). Using OPH, a vehicle is either assigned to be donating or receiving, and hence cannot do both in a single iteration. Even with multiple iterations, the toy example would require other vehicles in the schedule that can receive orders 2 and 4 while maintaining individual rationality. 

\subsection{Synthetic Data Benchmark}
First, we recreate the synthetic data benchmark used in \cite{de_jonge_heuristic_2021}. As explained in \cite{de_jonge_heuristic_2021}, this data set is generated from the Li \& Lim data set\cite{inproceedings}. The steps to generate each problem is as follows:

\begin{enumerate}
    \item Pick 2 separate files ($A$ and $B$), representing orders from LSP1 and LSP2 respectively.
    \item Offset the coordinates in B by ($\delta_x$, $\delta_y$). This includes the depot.
    \item Merge the 2 sets of orders as $A + B (\delta_x, \delta_y)$
\end{enumerate}

The rationale for applying an offset is to generate a larger pool of data and pick the top 5 best configurations in terms of potential improvement per category in the Li \& Lim data. We generated 10 out of 15 of the same configurations shown in \cite{de_jonge_heuristic_2021} and run both OPH and GAT to obtain the results shown in \autoref{tab:syntheticdata}. OPH is run for 1 iteration while GAT is run for a maximum of 5 iterations.

Take note that our method requires revenue of orders to compute the profit for each LSP to use for the IR constraint. To make our problem equivalent to the benchmarks in \cite{de_jonge_heuristic_2021} where we minimise the travel cost only, we set the revenue of all orders to be 0.

\subsection{Real Data}
Next, we run OPH and GAT on our real-life data, which has a few differences. The real data uses orders in Singapore, and computes time and distance values via OSRM. Due to the nature of the orders and the size of Singapore, the time taken to travel between waypoints and the service time is smaller relative to the time windows, allowing routes to contain approximately 2 times as many orders as compared to routes in the synthetic data. Another significant difference is the number of LSPs involved in each problem. In our real-life data we have 5 LSPs, providing a total of 200 orders. 

Because we now have revenue data, the social welfare improvement results shown in \autoref{tab:realdata} reflect the percentage improvement in profits.

\subsection{Results Analysis}

The results in both \autoref{tab:syntheticdata} and \autoref{tab:realdata} show that GAT outperforms OPH consistently, while having a more consistent and lower time taken to solve. In \cite{de_jonge_heuristic_2021}, it is noted that OPH solve times can have very large variations depending on the data. This is due to the complexity of generating one-to-one exchanges in OPH, which have a $O(N^2 V)$ worst case time complexity (where $N$ is the total number of orders and $V$ the total number of vehicles), assuming the VRP solve time to be $O(1)$. The number of order packages can vary significantly from 600 to 1000+ and the pre-processing step to filter infeasible order package - vehicle pairs depend heavily on the time windows, service time, travel times, and volume of the orders. This issue is highlighted further in our real data results \autoref{tab:realdata}, as vehicles can contain a large number of orders, and orders can be easily inserted into routes while maintaining feasibility.

In comparison, the main computation in GAT when solving the 2-vehicle VRPs has a time complexity of $O(V^2)$. While the 2-vehicle VRPs being solved in GAT are more complex compared to the 1-vehicle VRPs solved in OPH, the solve time is not significantly higher. In many cases, the 2-vehicle VRP contains an empty vehicle from one of the LSPs or a vehicle with <5 orders, which makes the solve time close to the OPH 1-vehicle VRP solve time. Since the number of vehicles is stable, relative to the number of feasible order packages in GOT, the time taken to solve each problem is more consistent.

GAT can produce much more complex exchanges as compared to OPH, especially since OPH is only run over a single iteration. We note however that we can run OPH over multiple iterations and could produce comparable results with GAT at the cost of long computation time. We run a different set of experiments, generating mock data from the real data where each problem consists of 6 LSPs with a fleet of 10 vehicles each and a total of only 40 orders. This is an edge case where the number of vehicles is larger than the number of orders. Using this new data set, we run both OPH and GAT over at most 5 iterations and the results are shown in \autoref{tab:smalldata}. While OPH outperforms GAT in 3 out of the 5 test cases, the time taken to solve all test cases is still larger than GAT.

\newcolumntype{s}{>{\hsize=.4\hsize}X}

\begin{table*}[t!]
    \centering
    \caption{Performance of Give-And-Take (GAT) and Order Package Heuristic (OPH) on the synthetic data benchmark in terms of social welfare and running time. Social welfare is defined as the percentage improvement of the total distance from $Init$}

    \label{tab:syntheticdata}
    \begin{tabularx}{\textwidth}{{>{\centering}X}*{4}{>{\centering\arraybackslash}s}}
        \toprule
        Test Case   & OPH Soc. Welf.   & OPH Time(s)   & GAT Soc. Welf.   & GAT Time(s) \\
        \midrule
        LC1\textunderscore2\textunderscore2 + LC1\textunderscore2\textunderscore6 (42,-42) &11.82$\%$ &22.28 &\textbf{12.43}$\%$ &74.00 \\
        LC1\textunderscore2\textunderscore2 + LC1\textunderscore2\textunderscore7 (-32,-32) &7.84$\%$ &19.54 &\textbf{11.04}$\%$ &96.32 \\
        LC1\textunderscore2\textunderscore4 + LC1\textunderscore2\textunderscore7 (-30,0) &7.50$\%$ &130.98 &\textbf{16.72}$\%$ &141.40 \\
        LC1\textunderscore2\textunderscore4 + LC1\textunderscore2\textunderscore8 (-30,0) &6.37$\%$ &368.96 &\textbf{14.58}$\%$ &145.64 \\
        LC1\textunderscore2\textunderscore10 + LC1\textunderscore2\textunderscore4 (30,0) &5.21$\%$ &650.31 &\textbf{18.44}$\%$ &155.34 \\
        LR1\textunderscore2\textunderscore3 + LR1\textunderscore2\textunderscore8 (0,30) &6.44$\%$ &442.28 &\textbf{19.76}$\%$ &213.21 \\
        LR1\textunderscore2\textunderscore5 + LR1\textunderscore2\textunderscore8 (0,30) &4.38$\%$ &209.56 &\textbf{12.10}$\%$ &243.50 \\
        LR1\textunderscore2\textunderscore8 + LR1\textunderscore2\textunderscore9 (0,-30) &4.34$\%$ &318.32 &\textbf{18.22}$\%$ &233.32 \\
        LR1\textunderscore2\textunderscore10 + LR1\textunderscore2\textunderscore3 (0,-30) &4.75$\%$ &288.30 &\textbf{16.92}$\%$ &170.44 \\
        LR1\textunderscore2\textunderscore10 + LR1\textunderscore2\textunderscore8 (0,30) &4.67$\%$ &553.00 &\textbf{18.63}$\%$ &248.72 \\
        \bottomrule
    \end{tabularx}
\end{table*}

\begin{table*}[t!]
    \centering
    \caption{Performance of Give-And-Take (GAT) and Order Package Heuristic (OPH) on our real world data in terms of social welfare and running time. Social welfare is defined as the percentage improvement of the sum of profits from $Init$}

    \label{tab:realdata}
    \begin{tabularx}{\textwidth}{*{5}{>{\centering\arraybackslash}X}}
        \toprule
        Test Case   & OPH Soc. Welf.   & OPH Time(s)   & GAT Soc. Welf.   & GAT Time(s) \\
        \midrule
        A &4.36$\%$ &538.62 &\textbf{14.32}$\%$ &270.77 \\
        B &5.23$\%$ &427.58 &\textbf{13.82}$\%$ &224.96 \\
        C &5.40$\%$ &325.55 &\textbf{12.65}$\%$ &235.70 \\
        D &6.34$\%$ &453.49 &\textbf{11.85}$\%$ &270.86 \\
        E &3.76$\%$ &666.15 &\textbf{12.25}$\%$ &274.30 \\
        \bottomrule
    \end{tabularx}
\end{table*}
\begin{table*}[t!]
    \centering
    \caption{Performance of Give-And-Take (GAT) and Order Package Heuristic (OPH) with multipe iterations on small mock-up data(6 LSPs, 40 orders) in terms of social welfare and running time. Social welfare is defined as the percentage improvement of the sum of profits from $Init$}

    \label{tab:smalldata}
    \begin{tabularx}{\textwidth}{*{7}{>{\centering\arraybackslash}X}}
        \toprule
        Test Case   & OPH Soc. Welf.   & OPH Time(s)   & GAT Soc. Welf.   & GAT Time(s) \\
        \midrule
        F &\textbf{41.31}$\%$ &165.75 &37.81$\%$ &30.79 \\
        G &29.27$\%$ &204.39 &\textbf{38.46}$\%$ &23.78 \\
        H &\textbf{29.11}$\%$ &111.68 &26.77$\%$ &24.69 \\
        I &15.32$\%$ &43.69 &\textbf{16.38}$\%$ &20.76 \\
        J &\textbf{17.36}$\%$ &183.22 &16.89$\%$ &29.06 \\
        \bottomrule
    \end{tabularx}
\end{table*}

\section{Conclusion and Future Works}
This paper presents a novel approach to the Collaborative Vehicle Routing Problem (CVRP), introducing the concept of Give-And-Take Exchange. By applying the principles of the Vehicle Routing Problem (VRP) to pairs of vehicles from different logistics companies, our approach optimizes the overall route while considering the constraints and objectives of both vehicles. This represents a significant departure from traditional CVRP solutions, which typically focus on finding exchange orders between vehicles.

While our work represents a significant advancement in the field of CVRP, there is still much work to be done. Future research could extend VRP for more than two vehicles and provide a substantial theorem to prove the efficiency of more than two vehicles. Additionally, further improvements to our algorithm could be explored, such as incorporating machine learning techniques to enhance the decision-making process.

\clearpage\clearpage

\bibliographystyle{named}
\bibliography{ijcai23}

\typeout{get arXiv to do 4 passes: Label(s) may have changed. Rerun}

\end{document}